\documentclass[aps,pra,floatfix,superscriptaddress,twocolumn,showpacs,10pt]{revtex4-1}
\usepackage{amssymb, amsmath,color,mciteplus,graphicx,subfigure}
\usepackage{calc,epsfig,epstopdf,color,mciteplus,bm,mathrsfs}
\usepackage{times}
\usepackage{float}
\usepackage{lipsum}

\begin{document}

\title{High-fidelity and Robust Geometric Quantum Gates that Outperform Dynamical Ones}

\author{Tao Chen}
\affiliation{Guangdong Provincial Key Laboratory of Quantum Engineering and Quantum Materials,
and School of Physics\\ and Telecommunication Engineering, South China Normal University, Guangzhou 510006, China}

\author{Zheng-Yuan Xue}\email{zyxue83@163.com}
\affiliation{Guangdong Provincial Key Laboratory of Quantum Engineering and Quantum Materials,
and School of Physics\\ and Telecommunication Engineering, South China Normal University, Guangzhou 510006, China}

\affiliation{Guangdong-Hong Kong Joint Laboratory of Quantum Matter, Frontier Research Institute for Physics,\\ South China Normal University, Guangzhou 510006, China}

\date{\today}

\begin{abstract}
  Geometric phase is a promising element to induce high-fidelity and robust quantum operations due to its built-in noise-resilience feature. Unfortunately, its practical applications are usually circumscribed by requiring complex interactions among multiple levels/qubits and the longer gate-time than the corresponding dynamical ones. 
  Here, we propose a general framework of geometric quantum computation with the integration of the time-optimal control technique, where the shortest smooth geometric path is found to realize accelerated geometric quantum gates, and thus greatly decreases the gate errors induced by both the decoherence effect and operational imperfections. Meanwhile, we faithfully implement our idea on a scalable platform of a two-dimensional superconducting transmon-qubit lattice, with simple and experimental accessible interactions. In addition, numerical simulations show that our implemented geometric gates possess higher fidelities and stronger robustness, which outperform the best performance of the corresponding dynamical ones. Therefore, our scheme provides a promising alternative way towards scalable fault-tolerant solid-state quantum computation.
\end{abstract}

\maketitle

\section{Introduction}

Based on fundamental quantum-mechanical principles, quantum computation (QC) can effectively deal with certain complex tasks that are hard for classical computers \cite{Nielson}, due to the intrinsic quantum parallelism. Therefore, various systems have been suggested for the physical implementation of QC, e.g., trapped ions \cite{ions}, cavity quantum electrodynamics (QED) \cite{cavity}, neutral atoms in optical lattices \cite{atom1,atom2}, etc. However, considering the requirements of large-scale integrability and flexibility, one of the promising candidates is the superconducting nanocircuit system \cite{jj1,jj2,jj3}, as it is compatible with modern ultrafast optoelectronics as well as with nanostructure fabrication and characterization. Besides, recent experimental advances in controlling the coherent evolution of quantum states \cite{DRAGexperiment1, TC5,TC6, QS2019} in larger lattices confirm the superconducting circuit as an interesting candidate with which to implement scalable QC, which requires at least a two-dimensional (2D) lattice of coupled qubits. Currently, control-induced crosstalk (frequency drift) among adjacent qubits in large qubit lattices is a main source of error for the implementation of quantum gates in that environment. Meanwhile, due to the inevitable interaction with the surrounding environment, the coherence of the quantum system is very fragile. Therefore, the question of how to suppress the effects of quantum operational imperfections and decoherence is the main challenge in realizing scalable QC.

To suppress quantum operational imperfections, quantum gates induced by geometric phases are promising \cite{zanardi}, due to their built-in noise-resilience features. Explicitly, geometric QC (GQC) \cite{zanardi,AGQC1,Duan} has been proposed through the use of adiabatic geometric phases. However, due to the long gate time required by the adiabatic condition, the decoherence effect causes considerable gate infidelity. To overcome such limitations, nonadiabatic GQC has been proposed to achieve high-fidelity quantum gates based on both Abelian \cite{wxb,ZSL1,UGQCZhu,NGQC,Chen2018} and non-Abelian geometric phases \cite{EJ,TongDM,Liu18}. Remarkably, experimental demonstrations of elementary gates for GQC have been made on various systems \cite{exp1, exp2, DuJ, Abdumalikov35, Feng39, Zu41, xuy37,yan2019, zhu2019, Xu2019}. However, due to the need for additional auxiliary energy levels beyond the qubit states and/or additional auxiliary coupling elements, the implementations of high-fidelity universal geometric gates are still experimentally challenging. Meanwhile, as for the decoherence effect, the time needed for a nonadiabatic geometric gate is still greater than for the corresponding dynamical one, leading to more gate infidelities: thus, this is the other major drawback of GQC. Finally, geometric evolution is usually based on  nonsmooth evolution paths, which further weakens the intrinsic robustness of the gate .

Here, we propose a general framework for nonadiabatic GQC in a simple setup, which integrates the time-optimal control (TOC) technique \cite{TO1,TO2} to find the shortest geometric evolution path for accelerated geometric quantum gates with smooth evolution paths, and thus can overcome the above-mentioned disadvantages of GQC perfectly. Meanwhile, our protocol can be realized on a 2D square lattice of superconducting qubits, where adjacent qubits are capacitively coupled, without increasing the circuit complexity by adding any additional auxiliary levels and qubits. In addition, we only utilize experimentally accessible two-body interaction by the parametrically tunable coupling \cite{TC6,TC5,TH1,TH2,TCExpYY}. Our numerical simulations show that, compared with dynamical gates, our implemented geometric gates can perform with higher gate fidelities and stronger robustness. Therefore, our protocol provides a promising route towards scalable fault-tolerant solid-state quantum computation.


\section{General framework}

\begin{figure}[tbp]
  \centering
  \includegraphics[width=0.95\linewidth]{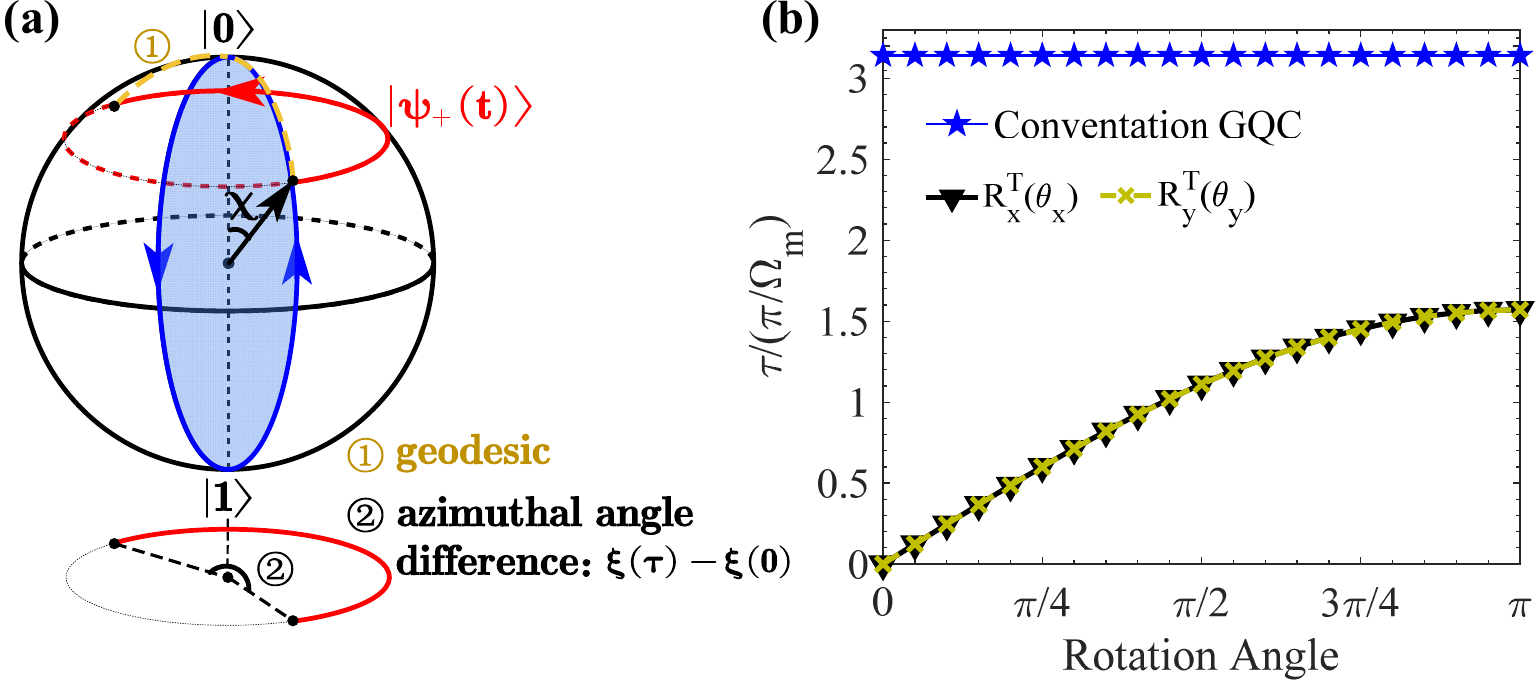}
  \caption{A comparison between TOC and conventional GQC. (a) A geometric illustration of the evolution paths of the TOC (red line) and conventional (blue line) geometric gates in a Bloch sphere. (b) The X- and Y-axis-rotation gate time of TOC and conventional GQC as a function of rotation angles, with the same time-dependent pulse shape of $\Omega(t)=\Omega_\textrm{m} \sin(\pi t/\tau)$.} \label{Fig1}
\end{figure}

We first illustrate how to implement nonadiabatic evolution for a general two-state system. In the rotating framework with respect to the driving frequency, assuming hereafter that $\hbar=1$, a general Hamiltonian for a driven two-level system is
\begin{eqnarray}
\label{EqH}
\mathcal{H}(t)=\frac {1} {2}
\left(
\begin{array}{cccc}
 -\Delta(t)             & \Omega(t) e^{-i\phi(t)} \\
 \Omega(t) e^{i\phi(t)} & \Delta(t)
\end{array}
\right),
\end{eqnarray}
where the basis consists of a ground state $|0\rangle=(1,0)^\dag$ and an excited state $|1\rangle=(0,1)^\dag$; $\Omega(t)$ and $\phi(t)$ are the amplitude and phase of the driving microwave field, respectively; $\Delta(t)$ is the time-dependent detuning between the qubit transition frequency and the frequency of the microwave field. As is well known, there is a pair of state vectors, $\{|\Psi_{\pm}(t)\rangle\}$, that satisfy the Schr\"{o}dinger equation of $\mathcal{H}(t)$ and, in this basis, the time-evolution operator is
\begin{eqnarray}
\label{EqUo}
U(t)=\textbf{T}e^{i\int\mathcal{H}(t)\textrm{d}t}\!=\!\!\!\sum_{j\in\{+,-\}}\!\!\!|\Psi_{j}(t)\rangle\langle \Psi_{j}(0)|, \notag
\end{eqnarray}
where $\textbf{T}$ is the time-ordering operator.
Here, to induce our target geometric phases, we introduce a set of auxiliary basis $|\psi_j(t)\rangle$ defined by $|\psi_j(t)\rangle=e^{-if_j(t)}|\Psi_j(t)\rangle$ with $f_j(0)=0$, which does not need to satisfy the Schr\"{o}dinger equation. For the selection of the auxiliary basis vectors, we start from the dynamic Lewis-Riesenfeld invariant $I(t)$ of $\mathcal{H}(t)$, satisfying
$\partial I(t)/\partial t+i[\mathcal{H}(t),I(t)]=0$, which is \cite{IR1,IR2,IR3}
\begin{eqnarray}
\label{EqIV}
I(t)=
\frac {\mu}{2} \left(
\begin{array}{cccc}
 \cos\chi(t)             & \sin\chi(t) e^{-i\xi(t)} \\
 \sin\chi(t) e^{i\xi(t)} & -\cos\chi(t)
\end{array}
\right),
\end{eqnarray}
where $\dot{\xi}(t)=-\Delta(t)-\Omega(t)\cot\chi(t)\cos[\phi(t)-\xi(t)]$ and $\dot{\chi}(t)=\Omega(t)\sin[\phi(t)-\xi(t)]$, in which $\mu$ is an arbitrary constant. As a matter of fact, we can select its eigenvectors,
\begin{eqnarray}
\label{EqEC}
\left\{
\begin{array}{ll}
|\psi_+(t)\rangle=\cos{\frac {\chi(t)} {2}}|0\rangle+\sin{\frac {\chi(t)} {2}}e^{i\xi(t)}|1\rangle, \\
\ \ \\
|\psi_-(t)\rangle=\sin{\frac {\chi(t)} {2}}e^{-i\xi(t)}|0\rangle-\cos{\frac {\chi(t)} {2}}|1\rangle, \\
\end{array}
\right.
\end{eqnarray}
as a set of the auxiliary basis: their evolutionary details of them in a Bloch sphere are visualized by the time-dependent polar angle $\chi(t)$ and azimuthal angle $\xi(t)$, as shown in Fig. \ref{Fig1}(a). That is, by determining the target evolution path dominated by $\chi(t)$ and $\xi(t)$, the Hamiltonian parameters $\{\Omega(t),\Delta(t),\phi(t)\}$ can then be reverse engineered \cite{IR3}. Therefore, at a final time $\tau$, the two dressed states are $U(\tau)|\psi_\pm(0)\rangle=e^{\pm i\gamma}|\psi_\pm(\tau)\rangle$ with $\pm\gamma\!=\!f_\pm(\tau)$, and the corresponding evolution operator is
\begin{eqnarray}
\label{EqU}
U(\tau)=e^{i\gamma}|\psi_+(\tau)\rangle\langle \psi_+(0)|+e^{-i\gamma}|\psi_-(\tau)\rangle\langle \psi_-(0)|,
\end{eqnarray}
where $\gamma=\int^\tau_0\langle \psi_+(t)|(i\frac {\partial} {\partial t}-\mathcal{H}(t))|\psi_+(t)\rangle \textrm{d}t=\gamma_g+\gamma_d$ is the Lewis-Riesenfeld phase, in which
\begin{eqnarray}
\label{Eqd}
\gamma_d=\frac {1} {2}\int^\tau_0 [\dot{\xi}(t)\sin^2\chi(t)+\Delta(t)]/\cos\chi(t) \textrm{d}t
\end{eqnarray}
and
\begin{eqnarray}
\label{Eqg}
\gamma_g=-\frac {1} {2}\int^\tau_0 \dot{\xi}(t)\left[1-\cos\chi(t)\right] \textrm{d}t
\end{eqnarray}
are the dynamical and global geometric parts, respectively.
The geometric nature of $\gamma_g$ can be verified by the fact that it is half of the solid angle enclosed by the evolution path and the geodesic line that connects the initial point [$\chi(0), \xi(0)$] and the final point [$\chi(\tau), \xi(\tau)$] (for details, see Appendix A). However, the existence of the dynamical phase $\gamma_d$ will lead to the loss of the geometric noise-resilient feature: thus, effectively, the ways to deal with the dynamical phase include eliminating it or transforming it to hold geometric properties. Note that the elimination option requires multiple and/or nonsmooth evolution paths: it is the main culprit in limiting the geometric gate time and weakening the robustness of the geometric gate. Thus, here, we devote ourselves to applying the latter strategy, i.e., letting $\gamma_d$ to meet the form of $\gamma_d=\alpha_g+\ell\gamma_g$,  where $\ell$ is a gate-independent constant and $\alpha_g$ is a coefficient that is dependent only on the geometric feature of the quantum evolution path during gate operation and thus finally makes the phase $\gamma$ an unconventional geometric phase \cite{UGQCZhu,DuJ}.

Next, we focus on the dynamical phase $\gamma_d$. By further setting $\int_0^{\tau}[\dot{\xi}(t)+\Delta(t)]/ {\cos\chi(t)}\textrm{d}t=-\int_0^{\tau}\dot{\xi}(t)\textrm{d}t$, the dynamical phase will be $\gamma_d=[\xi(0)-\xi(\tau)]-\gamma_g$, thus satisfying the unconventional geometric condition \cite{DuJ}, with $\ell=-1$. Moreover, in this case, $\alpha_g=\xi(0)-\xi(\tau)$ is just the difference in azimuthal angle between the initial and final points and is independent of the polar angle $\chi(t)$, as shown in Fig. \ref{Fig1}(a), which also solely depends on the essential geometric feature of the overall evolution path. In addition, we retain the unchanged polar angle as $\chi(t)=\chi$, that is, making the dressed states evolve along the latitude line of the Bloch sphere. With these settings, the constraints for the dressed-state parameters reduce to
\begin{eqnarray}
\label{EqCon}
\xi(t)=\phi(t)+\pi, \quad \chi=\tan^{-1}(\Omega(t)/[\dot{\xi}(t)+\Delta(t)]),
\end{eqnarray}
and the resulting geometric evolution operator from Eq. (\ref{EqU}) is
\begin{eqnarray}
\label{EqUtau}
U(\tau)=\left(
\begin{array}{cccc}
(c_{\gamma'}+is_{\gamma'} c_\chi) e^{-i \xi_-}  & is_{\gamma'} s_\chi e^{-i \xi_+} \\
 is_{\gamma'} s_\chi e^{i\xi_+} & (c_{\gamma'}-is_{\gamma'} c_\chi) e^{i\xi_-}
\end{array}
\right),\ \ \ \ \ \
\end{eqnarray}
where $c_i=\cos i$, $s_i=\sin i$, and $\gamma'=\gamma+\xi_-=-\xi_-$ with $\xi_\pm=[\xi(\tau)\pm\xi(0)]/2$. In this way, the target control of the geometric X- and Y-axis rotation operations for arbitrary angles $[0,\pi]$ can all be done by determining $\xi_+=0$ and $\pi/2$ with the same $\xi_-=-\pi/2$, respectively, in a single evolution path. Note that, up to now, parameters $\xi(t)$ [or $\phi(t)$] and $\Omega(t)$ can take arbitrary shapes, provided that the boundary values of $\xi(t)$ are fixed to realize different rotation operations.

\begin{figure}[tbp]
  \centering
  \includegraphics[width=\linewidth]{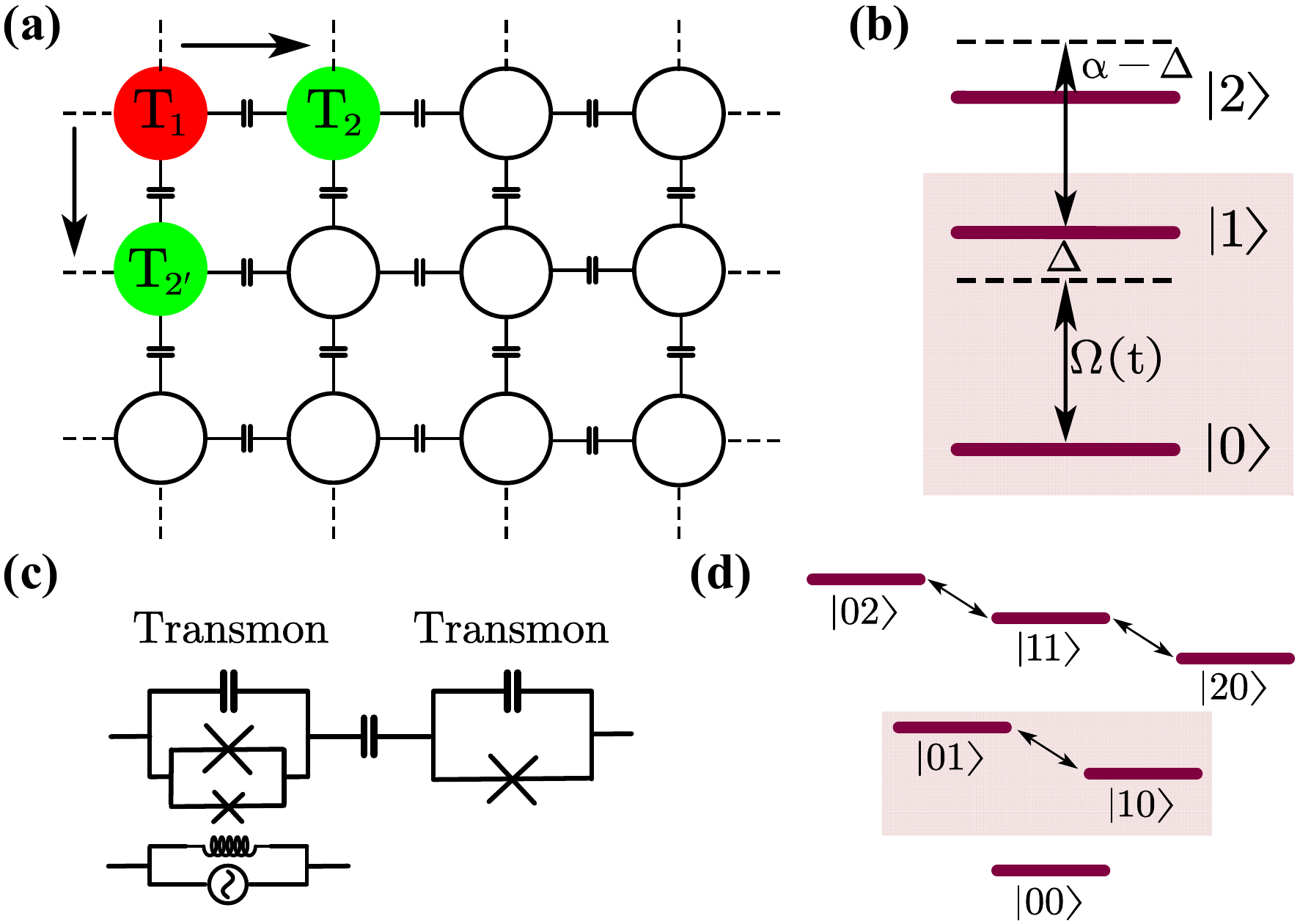}
  \caption{An illustration of our scheme. (a) A scalable 2D qubit lattice consisting of capacitively coupled superconducting transmons. (b) The energy levels and coupling configuration of a transmon, where an external microwave field is meant to couple the two lowest levels but will also stimulate transitions among the higher excited states, in a dispersive way. (c) The circuit details and (d) the energy spectrum of two capacitively coupled transmons, where the parametrically tunable coupling within the single-excitation subspace $\{|01\rangle, |10\rangle\}$ can be used to implement two-qubit geometric gates with TOC.} \label{Fig2}
\end{figure}

Furthermore, to pursue higher gate fidelity, we need to minimize the gate time to reduce the gate infidelity induced by the decoherence. Therefore, we can incorporate the TOC technique into our framework of GQC, i.e., by engineering the shape of $\phi(t)$ and $\Omega(t)$, to accelerate the target geometric gate. As for the quantum dynamics under the driving Hamiltonian $\mathcal{H}(t)$, the different selection of $\Omega(t)$ and $\phi(t)$ makes the quantum system evolve along different paths. The motivation of TOC is to find the path with the shortest time. And then, in the realistic physical implementation, the considered interaction Hamiltonian as $\mathcal{H}_c(t)=\frac {1} {2}\Omega(t)[\cos\phi(t)\sigma_x+ \sin\phi(t)\sigma_y]$ in Eq. (\ref{EqH}) needs to satisfy the following certain constraints: (i) the driving amplitude of the microwave field cannot be infinite, i.e., $f_1[\mathcal{H}_c(t)]=\frac{1}{2}[\textrm{Tr}(\mathcal{H}_c(t)^2)-\frac{1}{2}\Omega^2(t)]=0$, and (ii) the form of the interaction Hamiltonian $\mathcal{H}_c(t)$ cannot be arbitrary, i.e., $f_2[\mathcal{H}_c(t)]=\textrm{Tr}(\mathcal{H}_c(t) \sigma_z)=0$, in which $\sigma_{x,y,z}$ are the Pauli operators in the computational subspace $\{|0\rangle, |1\rangle\}$. Then, by solving the quantum brachistochrone equation \cite{QBE}
$\partial\textrm{F}/ \partial t=-i[\mathcal{H}(t),\textrm{F}]$, with $\textrm{F}=\partial (\sum_{j=1,2}\lambda_jf_j[\mathcal{H}_c(t)])/\partial \mathcal{H}(t)=\lambda_1\mathcal{H}_c(t)+\lambda_2 \sigma_z$, in which $\lambda_j$ is the Lagrange multiplier, we can determine the restricted parameter as follows:
\begin{eqnarray}
\label{restriction}
\phi(t)=\phi_0+\phi_1(t),\quad \phi_1(t)=\int^t_0[C_0\Omega(t')-\Delta(t)]\textrm{d}t', \ \
\end{eqnarray}
by defining $\lambda_1=1/\Omega(t)$ and $\lambda_2=-C_{0}/2$, where $\Omega(t)$ can be an arbitrary pulse shape, and the coefficient $C_{0}$ is a constant that depends only on the type of target gate. It is worth emphasizing that the time-dependent $\Omega(t)$ allows the incorporation of the pulse-shaping technique into the gate construction, which is essential in eliminating various gate errors. However, the fastest gate needs to correspond to a square pulse shape, i.e., $\Omega(t)$ needs to be time independent. Without loss of generality, here we present our framework with a time-dependent $\Omega(t)$. Then, the time-optimal form of the Hamiltonian $\mathcal{H}(t)$ in Eq. (\ref{EqH}) can be determined to realize geometric gates, the resulting evolution path under which is shorter than for the corresponding conventional ones \cite{NGQC}, as shown in Fig. \ref{Fig1}(a).

\section{Universal single-qubit geometric gates}

\begin{figure}[tbp]
  \centering
  \includegraphics[width=0.95\linewidth]{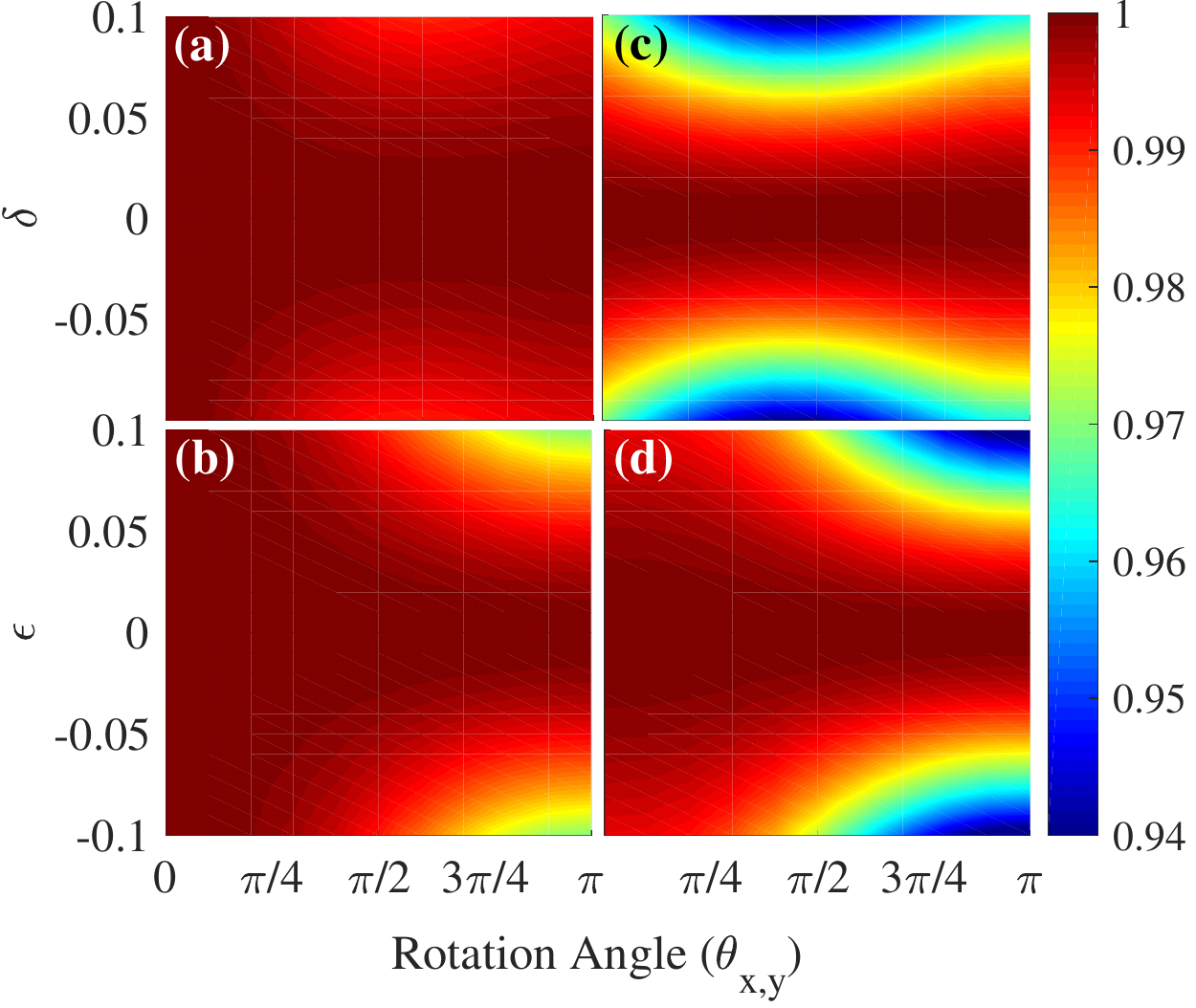}
  \caption{A comparison of the robustness for the X- and Y-axis rotation operations. The results of the gate fidelities versus the qubit frequency drift $\delta$ and the deviation $\epsilon$ of the driving amplitude for the (a),(b) geometric and (c),(d) dynamical X- and Y-axis rotation operations, respectively.} \label{Fig3}
\end{figure}

We now proceed to presenting our implementation of nonadiabatic GQC with the TOC technique on a 2D square superconducting transmon-qubit lattice, as shown in Fig. \ref{Fig2}(a), starting from a single transmon qubit, where the computational subspace $\{|0\rangle, |1\rangle\}$ consists of the ground and first excited states of the transmon. Conventionally, as shown in Fig. \ref{Fig2}(b), arbitrary control over the transmon qubit can be realized by applying a microwave field driven with the time-dependent amplitude $\Omega(t)$ and phase $\phi(t)$ on its two lowest levels, which leads to Eq. (\ref{EqH}) with a constant detuning $\Delta$. Note that a fixed $\Delta$ is more preferable experimentally, as it will not affect the qubit's coherent properties. To implement universal single-qubit geometric gates with TOC, we set the restrictions on the parameters in Eq. (\ref{restriction}) by defining $C_0=\cot(\theta/2)$. In this way, the geometric X- and Y-axis rotation operations with TOC, denoted as $R_x(\theta_x)$ and $R_y(\theta_y)$, for arbitrary angles $\theta_{x,y}\in[0,\pi]$ can all be realized by setting
\begin{eqnarray}
\label{MPA}
\phi_1(\tau_x)&=&-\pi,\ \ \theta=\theta_x,\ \ \phi_0=-\frac {\pi} {2};  \notag \\
\phi_1(\tau_y)&=&-\pi,\ \ \theta=\theta_y,\ \ \phi_0=0,
\end{eqnarray}
with the same $\phi_1(0)=0$ and the minimum pulse area of
\begin{eqnarray}
\label{MPA}
\frac {1} {2} \int^{\tau_{x,y}}_0 \Omega(t) \textrm{d}t=\frac {\pi} {2}/\sqrt{1+\cot^2\frac {\theta} {2}},
\end{eqnarray}
which are all less than $\pi$, required for conventional geometric operations \cite{EJ,NGQC}.
As an explicit proof, we take the simple pulse $\Omega(t)=\Omega_\textrm{m} \sin(\pi t/\tau)$ as an example: the time-acceleration results are shown in Fig. \ref{Fig1}(b).
In addition, we test the robustness of our geometric gates by utilizing the gate-fidelity formula $F^{\delta, \epsilon}=\textrm{Tr}(R^\dagger_{x,y}R^{\delta, \epsilon}_{x,y})/\textrm{Tr}(R^\dagger_{x,y}R_{x,y})$, in which $R^{\delta, \epsilon}_{x,y}$ are the affected rotation operations. Our simulation results in Fig. \ref{Fig3} show that, for both the qubit frequency drift $\delta$ and the deviation $\epsilon$ of the driving amplitude in the form of $\Delta+\delta\Omega_\textrm{m}$ and $\Omega(t)+\epsilon\Omega_\textrm{m}$, our implemented geometric gates possess better noise-resilient features than for the corresponding dynamical gates (for details, see Appendix B).

\begin{figure}[tbp]
  \centering
  \includegraphics[width=0.95\linewidth]{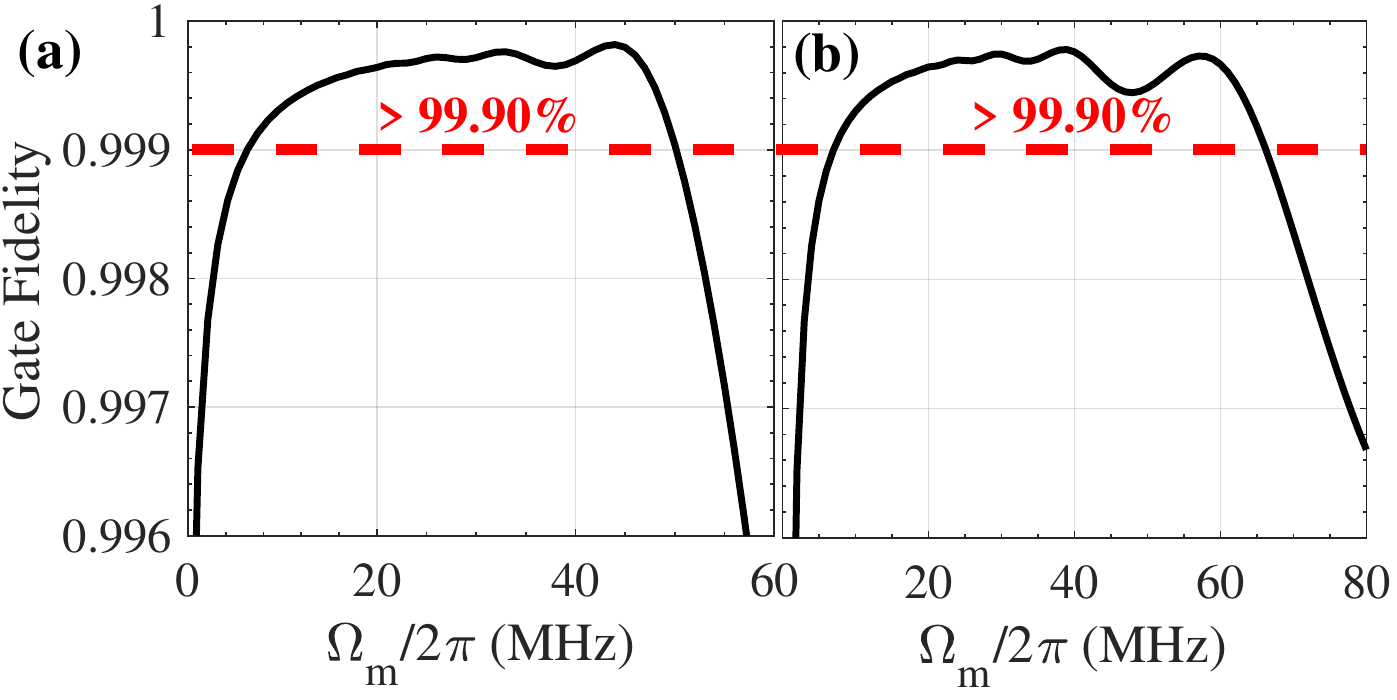}
  \caption{The performance of the single-qubit geometric gates. The results of the gate fidelities of the (a) $R_x(\pi/2)$ and (b) $R_y(-\pi/2)$ geometric operations as functions of the tunable parameter $\Omega_{\textrm{m}}$. 
  }\label{Fig4}
\end{figure}

In realistic physical implementations, due to the weak anharmonicity of the transmon qubit, a target driving on qubit states will also simultaneously stimulate the sequential transitions among the higher excited states, in a dispersive way. Targeting such an obstacle, we also apply the recent theoretical exploration of derivative removal via an adiabatic gate (DRAG) \cite{DRAG1,DRAG2} to suppress this type of leakage error, to obtain the precise qubit manipulation (for details, see Appendix C). To further analyze the performance of the single-qubit time-optimal geometric gates, we take the geometric operations $R_x(\pi/2)$ and $R_y(-\pi/2)$ as two typical examples. In our simulation, from the state-of-art experiment \cite{DRAGexperiment1}, we choose the relaxation and dephasing rates of the transmon to be identical, as $\kappa=\kappa^1_-=\kappa^1_z=2\pi\times 4$ kHz, and the anharmonicity to be $\alpha_1=2\pi\times 220$ MHz ( for the details of the simulation, see Appendix D). In Figs. \ref{Fig4}(a) and \ref{Fig4}(b), we plot the gate fidelities of $R_x(\pi/2)$ and $R_y(-\pi/2)$ as functions of tunable parameters, where we find that, when $\Omega_\textrm{m}=2\pi\times 45$ MHz with the corresponding restricted detuning parameter $\Delta\approx 2\pi\times 69$ MHz for $R_x(\pi/2)$, and $\Omega_\textrm{m}=2\pi\times 40$ MHz with $\Delta\approx 2\pi\times 11$ MHz for $R_y(-\pi/2)$, the fidelities of these two gates are both close to $99.98\%$, which compares favorably with the best performance of the reported experiments for the same type of gate  \cite{DRAGexperiment1}. 

\section{Nontrivial two-qubit geometric gate}

We next work on implementing the nontrivial two-qubit geometric gate with the TOC technique on the 2D square superconducting transmon-qubit lattice in Fig. \ref{Fig2}(a). For the capacitively coupled qubit lattice, the coupling strength of two adjacent transmons, e.g., $\textrm{T}_1$ and $\textrm{T}_2$ in the same row (or $\textrm{T}_1$ and $\textrm{T}_{2'}$ in the same column) is fixed. Meanwhile, the frequency difference of the two adjacent transmons, $\Delta_1=\omega_2-\omega_1$, is also generally hard to adjust, so that it is difficult to achieve resonant coupling and/or off-resonant coupling without changing a qubit frequency to deviate from its optimal working point. To deal with these difficulties, here we introduce an additional qubit-frequency driving for the transmon $\textrm{T}_1$, which can be experimentally realized by a longitudinal driving field, in the form of $\varepsilon(t)=\dot{F}(t)$ \cite{TCExpYY}, where $F(t)=\beta\sin[\nu t+\varphi(t)]$, with $\nu$ and $\varphi(t)$ indicating the frequency and phase of the longitudinal field, respectively: the circuit details are shown in Fig. \ref{Fig2}(c). Moving to the interaction picture, the coupling Hamiltonian reads
\begin{eqnarray}
\label{EqH12}
\mathcal{H}_{12}(t)&=& g_{_{12}} \left\{ |01\rangle_{_{12}}\langle 10|e^{i\Delta_1 t}+\sqrt{2}|11\rangle_{_{12}}\langle 20|e^{i(\Delta_1+\alpha_1) t}\right. \notag \\
&&+\left.\sqrt{2}|02\rangle_{_{12}}\langle 11|e^{i(\Delta_1-\alpha_2) t}\right\}e^{-i \beta\sin[\nu t+\varphi(t)]}\notag\\
&& +\mathrm{H.c.},
\end{eqnarray}
where $g_{_{12}}$ is the coupling strength between transmons $\textrm{T}_1$ and $\textrm{T}_2$ and $\alpha_j$ is the intrinsic anharmonicity of the transmon $\textrm{T}_j$. Utilizing the Jacobi-Anger identity and then neglecting the high-order oscillating terms, we find that the parametrically tunable coupling in the single-excitation subspace $\{|01\rangle_{_{12}}, |10\rangle_{_{12}}\}$ and the two-excitation subspace $\{|02\rangle_{_{12}}, |11\rangle_{_{12}}, |20\rangle_{_{12}}\}$ can all be selectively addressed by adjusting the frequency $\nu$. The corresponding energy spectra of these two capacitively coupled transmons, $\textrm{T}_1$ and $\textrm{T}_2$, are shown in Fig. \ref{Fig2}(d).

\begin{figure}[tbp]
  \centering
  \includegraphics[width=0.8\linewidth]{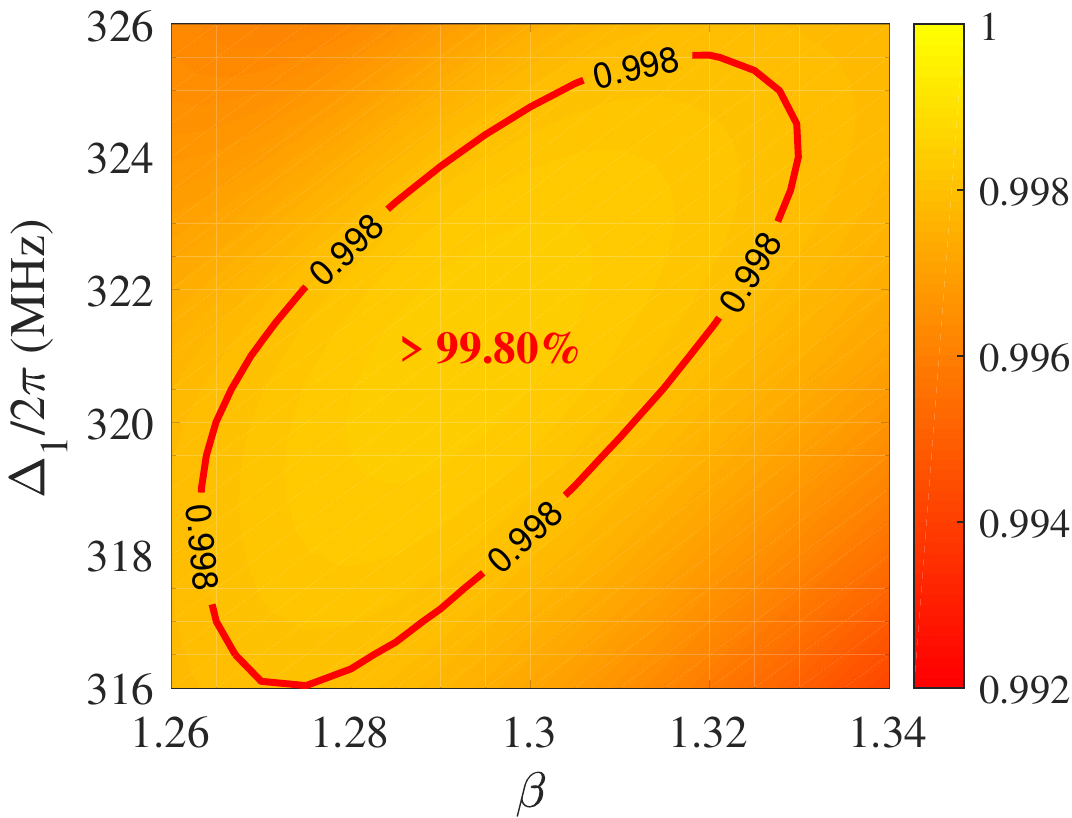}
  \caption{The performance of the implemented time-optimal two-qubit geometric gate in terms of the gate fidelities as functions of tunable parameters. 
  }\label{Fig5}
\end{figure}

However, the use of the interaction of high-excitation subspaces tends to cause more decoherence factors than that of the single-excitation subspace. Thus, here, we purposely pick the interactions of the single-excitation subspace. By setting the frequency $\nu$ to satisfy $\Delta_t=\nu-\Delta_1$, with $|\Delta_t|\ll\{\Delta_1, \nu\}$, and applying the unitary transformation, we can obtain the effective Hamiltonian as
\begin{eqnarray}
\label{EqH22}
\mathcal{H}_{t}(t)=\frac {1} {2}
\left(
\begin{array}{cccc}
 -\Delta_t                  & g'_{_{12}} e^{-i\varphi(t)} \\
 g'_{_{12}} e^{i\varphi(t)}& \Delta_t
\end{array}
\right),
\end{eqnarray}
in the single-excitation subspace $\{|01\rangle_{12},|10\rangle_{12}\}$, where the effective coupling strength $g'_{_{12}}=2J_1(\beta)g_{_{12}}$, in which $J_1(\beta)$ is a Bessel function of the first kind. As for the leakage from the computational basis $|11\rangle_{12}$ to the higher excitation subspaces, we can further restrain it by optimizing the system parameters. In the same way, in the above equivalent two-level system, to achieve integration with the TOC technique our restricted result in Eq. (\ref{restriction}) is denoted by $\varphi(t)=\varphi_0+\varphi_1(t)$, based on the effective square-pulse shape $g'_{_{12}}$, in which $\dot{\varphi_1}(t)=\eta$ is a constant that depends only on the type of target gate. Thus, within the two-qubit subspace $\{|00\rangle_{12},|01\rangle_{12},|10\rangle_{12},|11\rangle_{12}\}$, the final evolution operator
\begin{eqnarray}
\label{U2}
U_{2}(T)=
\left(
\begin{array}{cccc}
 1 & 0 & 0 & 0 \\
 0 & -\cos\frac {\vartheta} {2} & \sin\frac {\vartheta} {2}e^{-i\varphi_0} & 0 \\
 0 & -\sin\frac {\vartheta} {2}e^{i\varphi_0} & -\cos\frac {\vartheta} {2} & 0 \\
 0 & 0 & 0 & 1
\end{array}
\right),
\end{eqnarray}
with the minimal time cost of $T=T_0/\left[2\sqrt{1+\cot^2 (\vartheta/2)}\right]$ can be obtained by determining $\varphi_1(T)=-\pi$. At this point, a nontrivial two-qubit time-optimal geometric gate can be realized. Obviously, the gate time $T$ is also faster than the corresponding conventional geometric operation, the time being $T_0=2\pi/g'_{_{12}}$ \cite{EJ,NGQC}. We next take the two-qubit geometric gate with $\vartheta= \pi/2$ and $\varphi_0=\pi/2$ as an typical example to fully evaluate its gate performance (for details of the simulation, see Appendix D). Realistically, we choose the coupling strength of the two adjacent transmons as $g_{_{12}}=2\pi\times 8$ MHz, the anharmonicity of the second transmon as $\alpha_2=2\pi\times 180$ MHz, and the relaxation and dephasing rates of the transmon to be identical, as $\kappa=\kappa^1_-=\kappa^1_z=\kappa^2_-=\kappa^2_z=2\pi\times 4$ kHz. However, from the energy spectra of the two capacitively coupled transmons $\textrm{T}_1$ and $\textrm{T}_2$ as shown in Fig. \ref{Fig2}(d), we find that the effect of leakage into the noncomputational subspace $\{|02\rangle_{12},|20\rangle_{12}\}$ cannot be negligible completely \cite{TC6}. To avoid this type of leakage error as much as possible, it is necessary to optimize the qubit parameters to obtain a parameter range in which a high-fidelity two-qubit geometric gate can be achieved. As shown in Fig. \ref{Fig5}, we find an elliptical regime, within which a two-qubit gate with a fidelity higher than 99.80\% can be realized. In particular, the numerical results show that when $\Delta_1= 2\pi\times320$ MHz, $\nu=2\pi\times 340$ MHz, and $\beta\simeq 1.3$, our two-qubit geometric gate fidelity can be as high as 99.84\%, which compares favorably with the best performance of the currently reported experiments. 

\section{Discussion and conclusion}

In summary, we proposed a practical implementation of GQC in a simple experimental setup, which only utilizes an experimentally accessible two-body interaction and avoids the introduction of additional auxiliary energy levels beyond the qubit states and additional auxiliary coupling elements. Meanwhile, our scheme is robust against the main gate error sources and is less affected by the decoherence effect. 
As the required interaction in Eq. (\ref{EqH22}) is the same as that for the single-excitation subspace of exchange-coupled spin systems, our scheme can be readily extended to these systems, e.g., quantum dots, cavity QED, trapped ions, etc. Therefore, our implementation uses only existing experimental technologies to remedy the main drawbacks of GQC, leading to an ultrahigh gate fidelity that compares favorably with the best performance of the current reported experiments and thus makes it a promising strategy on the route towards robust and scalable solid-state QC.

\begin{acknowledgements}
We thank B.-J. Liu for helpful discussions. This work was supported by the Key-Area Research and Development Program of GuangDong Province (Grant No. 2018B030326001), the National Natural Science Foundation of China (Grant No. 11874156), the National Key R\&D Program of China (Grant No. 2016YFA0301803), Science and Technology Program of Guangzhou (Grant No. 2019050001), and the Innovation Project of Graduate School of SCNU (Grant No. 2019LKXM006).
\end{acknowledgements}

\appendix

\section{Derivation of the Global geometric phase $\gamma_g$}

Let us first consider the dressed state $|\psi_+(t)\rangle$ to examine its geometric nature of the evolution process. For our noncyclic evolution path $C_1$, the total relative phase between the initial state $|\psi_+(0)\rangle$ and the final state $U(\tau)|\psi_+(0)\rangle$ (i.e., $e^{i\gamma}|\psi_+(\tau)\rangle$) can be defined as $\gamma_t\!\!=\!\!\arg\langle \psi_+(0)|U(\tau)|\psi_+(0)\rangle$. Thus, following Refs. \cite{Pphase1,Pphase2,Pphase3,Pphase4}, the geometric Pancharatnam phase in path $C_1$ can be determined as
\begin{eqnarray}
\label{EqPPhase}
\gamma_p(C_1)=&&\gamma_t-\gamma_d \notag \\
=&&\arg\langle \psi_+(0)|U(\tau)|\psi_+(0)\rangle \notag \\
&& +\!\int^\tau_0\!\!  \langle\psi_+(t)|\mathcal{H}(t)|\psi_+(t)\rangle \textrm{d}t,
\end{eqnarray}
where $\gamma_d$ is the corresponding dynamical phase. In addition, under a gauge transformation $|\psi_+(t)\rangle\!\!\rightarrow\!\!|\psi'_+(t)\rangle\!=\!e^{i\varsigma(t)}|\psi_+(t)\rangle$, we have
\begin{eqnarray}
\label{EqPPhase}
\gamma'_p(C_1)
=&&\arg\langle \psi'_+(0)|U(\tau)|\psi'_+(0)\rangle\notag \\
&&+\int^\tau_0\!\!  \langle\psi_+(t)|\mathcal{H}(t)|\psi_+(t)\rangle \textrm{d}t \notag \\
=&&\arg\langle \psi_+(0)|U(\tau)|\psi_+(0)\rangle+\varsigma(\tau)-\varsigma(0) \notag \\
&&+\int^\tau_0\!\!  \langle\psi_+(t)|\left(\mathcal{H}(t)-\dot{\varsigma}(t)\right)|\psi_+(t)\rangle \textrm{d}t \notag \\
=&&\gamma_p(C_1);
\end{eqnarray}
i.e., the Pancharatnam phase $\gamma_p(C_1)$ is gauge invariant. Following the geodesic rule \cite{Pphase1,Pphase4}, for the noncyclic geometric evolution, it is assumed that there is a geodesic line that connects the initial [$\chi(0), \xi(0)$] and final [$\chi(\tau), \xi(\tau)$] points of the actual evolution path, so that the actual evolution path $C_1$ plus the geodesic line $C_2$ can form a closed path $C$. According to the similar formula above, we can solve for the accumulated Pancharatnam phase $\gamma_p(C_2)$ under the geodesic line, which is also gauge invariant. Thus, the overall Pancharatnam phase is
\begin{eqnarray}
\label{EqPPhaseC}
\gamma_p(C)=\gamma_p(C_1)\!+\!\gamma_p(C_2)\!=\!\frac {1} {2}\int_{C} \sin\chi \textrm{d}\chi\textrm{d}\xi,
\end{eqnarray}
which just represents half of the solid angle enclosed by the evolution path and the geodesic line. As $\textrm{d}\xi=0$ on the additional geodesic line $C_2$, the result of $\gamma_p(C)$ will reduce to
$$\gamma_p(C)=-\frac {1} {2}\int^\tau_0 \dot{\xi}(t)\left[1-\cos\chi(t)\right] \textrm{d}t=\gamma_g$$ which is the global geometric phase in Eq. (\ref{Eqg}) of the main text; thus it also has the geometric nature. A similar discussion is applicable for the state  $|\psi_-(t)\rangle$.

\section{Comparison of gate robustness}

The pursuit of high-fidelity and strongly robust quantum gates in the previous geometric schemes is usually circumscribed by complex interactions among multiple levels or qubits and the longer time required than for the dynamical counterpart. Therefore, it is desirable to prove that our geometric gate has a stronger robustness than the dynamical one, which can be realized only by changing $\phi(t)$ as a constant, i.e., $\phi(t)=\phi_0$, to ensure the geometric phase $\gamma_g=0$. In particular, to ensure the fairness of our gate-robustness comparison, here we define the pulse shape of $\Omega(t)$ to be the same as that of geometric gates. The resulting dynamical evolution operator can be obtained as
\begin{eqnarray}
\label{EqUd}
&&U_d(\lambda,\theta_d,\phi_0) \notag \\
&&\quad \quad =\cos \frac {\lambda} {2}-i\sin\frac {\lambda} {2}
\left(
\begin{array}{cccc}
 -\cos\frac {\theta_d} {2}          & \sin\frac {\theta_d} {2} e^{-i\phi_0} \\
 \sin\frac {\theta_d} {2} e^{i\phi_0} & \cos\frac {\theta_d} {2}
\end{array}
\right),\quad
\end{eqnarray}
with $\lambda=\int_0^{\tau}\sqrt{\Omega^2(t)+\Delta^2(t)}\textrm{d}t$ and $\theta_d=2\tan^{-1}[\Omega(t)/\Delta(t)]$. In this way, arbitrary dynamical X- and Y-axis rotation operations can be all realized by $U_d(\pi,2\pi,0)U_d(\pi,\theta_x,-\pi/2)$ and $U_d(\pi,2\pi,0)U_d(\pi,\theta_y,0)$. Unlike the implementation of our geometric gate, in the dynamical case, one needs to restrict the phase variable $\phi(t)$ to be time independent in order to realize a target gate: thus there is no additional degree of freedom to combine with TOC.

\section{DRAG correction}

In the realistic physical implementation, due to the weak anharmonicity of transmons, when our target microwave field is applied to the two lowest levels of transmon, it will also simultaneously induce the sequential transitions among the higher excited states, resulting in the leakage error. Thus, to achieve independent manipulation of qubit states, we apply the recent theoretical exploration of DRAG \cite{DRAG1,DRAG2} to suppress this type of leakage error. Here, we consider the influence of the third energy level, which is the main leakage source of our qubit states. To this end, the Hamiltonian describing a single-qubit system can be written as
\begin{eqnarray}
\label{EqHS}
\mathcal{H}_1(t)=\frac {1} {2}\textbf{B}(t)\cdot \textbf{S}-\alpha_1 |2\rangle\langle 2|,
\end{eqnarray}
where $\alpha_1$ is the intrinsic anharmonicity of the target transmon, $\textbf{B}(t)=\textbf{B}_0(t)+\textbf{B}_d(t)$ is the vector of the total microwave field including the original and additional DRAG-correcting microwave fields, i.e.,
\begin{eqnarray}
\label{EqB0}
\textbf{B}_0(t)&=&(B_x,B_y,B_z)              \notag \\
&=&(\Omega(t)\cos(\phi_0+\phi_1(t)), \Omega(t)\sin(\phi_0+\phi_1(t)), -\Delta), \notag \\
\textbf{B}_d(t)&=&(B_{d;x},B_{d;y},B_{d;z})              \notag \\
&=&-\frac {1} {2\alpha_1}(-\dot{B}_y+B_zB_x, \dot{B}_x+B_zB_y, 0), \notag
\end{eqnarray}
respectively, and the operator vector $\textbf{S}$ is given by
\begin{eqnarray}
\label{S}
S_x&=&\sum_{m=0,1}\sqrt{m+1}(|m+1\rangle\langle m|+|m\rangle\langle m+1|), \notag \\
S_y&=&\sum_{m=0,1}\sqrt{m+1}(i|m+1\rangle\langle m|-i|m\rangle\langle m+1|),\notag \\
S_z&=&\sum_{m=0,1,2}(1-2m)|m\rangle\langle m|. \notag
\end{eqnarray}
Meanwhile, through numerical simulation, we find that, for all the implemented single-qubit geometric gates, their infidelities caused by leakage to the third energy level are all less than $0.01\%$, which is almost negligible, thus confirming that it is feasible for the DRAG correction in our simulation.

\section{Master-equation simulation}

In practical physical implementations, the performance of our implemented geometric gate is inevitably limited by the decoherence effect of the target qubit system. In addition, to quantitatively evaluate the validity of the final effective Hamiltonian, all of our simulations hereafter are based on the original interaction Hamiltonian without any approximation. Therefore, here we consider the effects of decoherence and the high-order oscillating terms. The quantum dynamics can be simulated by the Lindblad master equation,
\begin{eqnarray}
\label{EqMaster}
\dot\rho_{_n}&=& -i[\mathcal{H}_d(t), \rho_{_n}]+\sum_{i=1}^n \left\{\frac {\kappa^i_-} {2}\mathscr{L}(|0\rangle_i \langle 1|+\sqrt{2}|1\rangle_i \langle 2|) \right. \notag \\
&&+\left. \frac {\kappa^i_z} {2}\mathscr{L}(|1\rangle_i \langle 1|+2|2\rangle_i \langle 2|) \right\},
\end{eqnarray}
where $\rho_{_n}$ is the density matrix of the quantum system under consideration, $\mathscr{L}(\mathcal{A})=2\mathcal{A}\rho_{_n}
\mathcal{A}^\dagger-\mathcal{A}^\dagger \mathcal{A} \rho_{_n} -\rho_{_n} \mathcal{A}^\dagger \mathcal{A}$ is the Lindblad operator for operator $\mathcal{A}$, and $\kappa^i_-$, $\kappa^i_z$ are the relaxation and dephasing rates of the $i$th transmon, respectively. For the cases of a single qubit and two coupled qubits, the forms of the driving Hamiltonians are expressed as $\mathcal{H}_d(t)=\mathcal{H}_1(t)$ with $n=1$ and $\mathcal{H}_d(t)=\mathcal{H}_{12}(t)$ with $n=2$, respectively.

To fully evaluate our implemented geometric gates, for the general initial state of the single qubit, $|\psi_1\rangle=\cos\theta_1|0\rangle+\sin\theta_1|1\rangle$, in which $|\psi_{f_{k=x,y}}\rangle=U(\tau_{k})|\psi_1\rangle$ is the ideal final state, we can define the single-qubit gate fidelity as $F_k^G=\frac {1} {2\pi}\int_0^{2\pi} \langle \psi_{f_k}|\rho_1|\psi_{f_k}\rangle \textrm{d}\theta_1$, where the integration is done numerically for 1001 input states, with $\theta_1$ being uniformly distributed within $[0, 2\pi]$, and $\rho_1$ is the numerically simulated density matrix of the qubit system. Furthermore, in the same way, in the two-qubit case, for a general initial state of the two qubits according to $|\psi_2\rangle=(\cos\vartheta_1|0\rangle+\sin\vartheta_1|1\rangle)\otimes
(\cos\vartheta_2|0\rangle+\sin\vartheta_2|1\rangle)$, in which $|\psi_{f_{U_2}}\rangle=U_2(T)|\psi_2\rangle$ is the ideal final state, we can also define the two-qubit gate fidelity as
\begin{eqnarray}
\label{Fidelity}
F^G_{U_2}=\frac {1} {4\pi^2}\int_0^{2\pi} \int_0^{2\pi} \langle \psi_{f_{U_2}}|\rho_2|\psi_{f_{U_2}}\rangle \textrm{d}\vartheta_1\textrm{d}\vartheta_2, \quad \quad
\end{eqnarray}
with the integration done numerically for 10001 input states, with $\vartheta_1$ and $\vartheta_2$ uniformly distributed over $[0, 2\pi]$.

\end{document}